# Photon data shed new light upon the GDR spreading width in heavy nuclei.

A.R. Junghans[1], G. Rusev[1,*], R. Schwengner[1], A. Wagner[1] and E. Grosse[1,2].

[1]*Institute for Radiation Physics, Forschungszentrum Dresden-Rossendorf, 01314 Dresden, Germany*
[2]*Institute for Nuclear and Particle Physics, Technische Universität Dresden, 01062 Dresden, Germany*

Received: Sept. 14, 2008
Received in revised form: Sept. 26, 2008

Supported by the DFG under Contract Nos. DFG-Gr1674/1-2 and DFG-Do466/1-2.

E-mail address: e.grosse@fzd.de

[*] Present address: Physics Dept., Duke University,
Durham NC 27708, USA.

**Abstract**
A global study of the electric dipole strength in and below the isovector giant dipole resonance (GDR) is presented for mass numbers A>80. It relies on the recently established remarkably good match between data for the nuclear photo effect to novel photon scattering data covering the region below the neutron emission threshold as well as by average resonance neutron capture (ARC). From the wide energy coverage of these data the correlation of the GDR spreading width with energy can be studied with remarkable accuracy. A clear sensitivity to details of the nuclear shape, i.e. the β- and γ-deformations, is demonstrated. Based hereon a new parameterization of the energy dependence of the nuclear electric-dipole strength is proposed which – with only two new parameters – allows to describe the dipole strength in all heavy nuclei with A>80. Although it differs significantly from previous parameterizations it holds for spherical, transitional, triaxial and well deformed nuclei. The GDR spreading width depends in a regular way on the respective resonance energy, but it is independent of the photon energy.

Key properties of nuclei are their mass and shape as well as their response to electromagnetic radiation. Photo-nuclear processes were among the first nuclear reactions studied [1] and their appreciable strength has triggered the conclusion [2] that they are likely to play an important role for the cosmic nucleosynthesis: In the intense photon flux during high temperature cosmic scenarios particle emission thresholds are reached leading to the photo-disintegration of previously formed nuclides. For a full assessment of photon induced processes the knowledge of the underlying smooth strength is similarly important as the "pygmy" structures observed in that energy range [3-5]. Finally, photon strength functions influence not only cosmic processes but they also are of importance for the detailed understanding of radiative neutron capture [6]: To analyze γ-spectra following capture the photon strength has to be known up to threshold. A detailed knowledge of neutron-induced processes is of practical importance for future systems dedicated to transmute nuclear waste as well as new concepts on nuclear reactors.

The electric dipole strength in heavy nuclei is mainly concentrated in the isovector giant dipole resonance (GDR). The centroid energy $E_0$ of the GDR is related to the symmetry-energy constant J and the surface-stiffness Q – as determined in a fit of the finite range droplet model (FRDM)[7] to the nuclear masses – with the effective nucleon mass m* as an additional parameter [8]. The energy-integrated dipole strength is determined by rather general quantum mechanical considerations leading to the Thomas-Reiche-Kuhn (TRK) sum rule [9]. Thus it is mainly the width of the GDR and its detailed shape which are of interest for further study. Very recently a covariant calculation based on density functional theory has resulted in a satisfactory description of the GDR for spherically symmetric nuclei [10]. Although equivalent calculations not restricted to spherical symmetry may be possible it is important to know how well our phenomenological understanding of nuclear masses and shapes can be extended to the dipole strength. This is the main subject of the comprehensive study presented here for heavy nuclei.

In this Letter electric dipole strength data for the GDR as well as for energies at and below the neutron threshold are compared for nuclei with A>80 to a new parameterization; this allows to shed new light upon the GDR width. Starting from the fact that the width due to particle escape is sufficiently small – as shown by respective calculations [11] – one has to quantify the other contributions to the apparent width of the GDR. On one hand these are due to the spreading into underlying complex configurations and on the other hand they are caused by a splitting induced by deformation of the nuclear shape. We use nuclear spectroscopy data to derive information on the shape and on its influence on the electric dipole strength. This allows us to extract the portion of the width caused by spreading and to derive a suitable parameterization for it. The basis for this is a Lorentzian parameterization [3] of the resonantly enhanced photo-absorption cross section $\langle\sigma_\gamma(E_\gamma)\rangle$ in the GDR and below – after averaging over the underlying many levels forming a quasi-continuum. Although not originating from the decay into the free vacuum, but rather from the spreading of the GDR strength into the nuclear quasi-continuum, the description of the dipole strength by Lorentzians has been proven [12] to be justified. Mainly from (γ,n) reactions [13, 14] a rather detailed experimental knowledge on the average absorption cross section <$\sigma_\gamma(E_\gamma)$> exists for many nuclei at energies well above the particle-separation energy $S_n$. In many nuclei near closed shells one-component Lorentz fits [14] to the GDR resulted in values for the product <$\sigma_{max}$>·Γ which considerably exceed the TRK-value 11.9/π·(NZ/A)MeV·fm²; in well deformed nuclei two Lorentzians have been used. Such fits [14] have neither resulted in a systematic scheme for the resonance widths Γ nor for the spreading. In contrast to this procedure, we apply a parameterization for the width and we rigorously require the integrated Lorentz curves to fulfil the TRK sum-rule. By comparing the result of such a calculation to data we avoid a fit, we demonstrate the role of the spreading width and we visualize eventually unaccounted effects.

In a first step, we make use of the abovementioned fact that the centroid energies $E_0$ of the GDR are well predicted by the FRDM [7, 8 (cf. Eq. 4.12)]:

$$E_0 = \frac{\hbar c}{R_0} \sqrt{\frac{8 \cdot J}{m^*} \cdot \frac{A^2}{4 \cdot N \cdot Z}} \cdot \left[1 + u - \varepsilon \cdot \frac{1+\varepsilon+3\cdot u}{1+\varepsilon+u}\right]^{-1/2}$$

$$\varepsilon = 0.0768 \qquad u = (1-\varepsilon) \cdot A^{-1/3} \cdot \frac{3 \cdot J}{Q} \qquad (1)$$

We use the standard fit to the nuclear masses by the FRDM [7] with $R_0 = 1.16$fm $\cdot A^{1/3}$, $J = 32.7$ MeV and $Q = 29.2$ MeV. The second term in the square root corrects for the approximation on Z/A made in [8]. From our comparison to the data for many heavy nuclei we get $m^* \cdot c^2 = 874$ MeV as the optimum value. A satisfactory expression for the resonance widths is not available in the FRDM [8] and a more refined description of the width has to be found. Because of the long lever arm data taken near the thresholds are especially sensitive to the energy dependence of the photon strength and thus the dipole spreading width. To correlate the dipole strength at energies below threshold to photo-nuclear data for the GDR the suggestive idea [3, 15, 16, 17] was pursued of extrapolating the Lorentzians to lower energies. The essential ingredient for such an extrapolation is the energy dependence of the resonance width, which is dominated by spreading. We extend the experimental basis by complementing the (γ,n)-data by results from photon scattering and from (n,γ)-results. To account for Porter-Thomas-fluctuations [15] the photon strength information has to be extracted from properly averaged measurements in both cases.

Photon scattering directly delivers the E1 strength at energies up to the thresholds [15-16, 18-22]. The cross section for photon scattering is identical to the photon absorption cross section as long as no particle emission or fission occurs. Theoretical arguments [15, 22] as well as data [16, 23] show that M1 and E2 excitations in heavy nuclei contribute only weakly to the photon absorption cross section between 5 MeV and the neutron binding energy $S_n$. It is thus justified to first concentrate on the influence of the E1-strength and then to test if it suffices to regard transitions starting from the ground state – as is usually done for smaller photon energies [24]. In case the experiment is performed in a bremsstrahlung continuum inelastic scattering cannot be directly identified. The necessary correction has been shown to be obtainable from Monte-Carlo methods [19, 20] which weakly depend on the energy dependence of the level density. Then the Axel-Brink rule [15] and a self consistency condition allow to extract from the data an electric-dipole strength- function with reasonable accuracy up to $S_n$ [19, 20]. Data obtained for many nuclei between $^{88}$Sr and $^{208}$Pb obtained at the Dresden Radiation Source ELBE have been analyzed that way [18-21]. The high photon flux and the strong background suppression combined with a photon detection system with favourable response [18] resulted in a rather high statistical significance of the data. Special care was taken to identify all strength by properly subtracting non-nuclear scattering. Porter-Thomas fluctuations were accounted for by averaging the raw data.

Photon strength information is also obtainable from the multi-step decay of neutron resonances [17, 25-27] or from the complex gamma decay following transfer reactions [28]. In both cases the absolute normalization has to come from independent data like neutron capture through resonances with accurately known photon widths. To reliably deduce from such data a photon strength and its energy dependence gamma decay branching ratios as well as spins and parities in the decay cascade have to be well known and thus the photon strengths determined such are likely to be hampered by inaccuracies [29, see section B]. To avoid possible ambiguities we use literature data only from ARC-experiments with sufficient averaging over many neutron resonances and with targets of ½⁻ and 3/2⁻, resulting in predominant E1 decay. As connective element between absorption and decay data a continuous electric-dipole strength-function $f_1(E_\gamma)$ was introduced [3]. It is derived from the average photon absorption cross section <$\sigma_\gamma(E_\gamma)$> but it can also be expressed by the ratio of the photon width <$\Gamma_{E1}$> to the level distance $D(E_\gamma, J^\pi=1^-)$ both averaged at the top of the electromagnetic transition. Relating these two processes one has for even-even nuclei [3]:

$$\frac{<\sigma_\gamma(E_\gamma)>}{3(\pi\hbar c)^2 \cdot E_\gamma} = f_1(E_\gamma) = \frac{<\Gamma_{E1}>}{E_\gamma^3 \cdot D} \qquad (2)$$

Applying the principle of detailed balance and the Axel-Brink rule [15] the strength 'upwards' from the ground state is identified to the 'downward' strength related to the average for E1 gamma decay of energy $E_\gamma$ between any two states.

In this Letter we investigate the energy dependence of the photon strength function $f_1(E_\gamma)$ to experimental information on eight nuclei of different shape and masses in the range from A=80 to A=240, selected by the extra requirement that reliable experimental information *above and below* threshold is available. In Figs.1-4 the situation is demonstrated for spherical, deformed, triaxial and transitional nuclei. In Fig.1 the results from using m*=874 MeV/c² and a single Lorentzian in Eq. (4) are displayed together with experimental data for the two spherical nuclei $^{88}$Sr and $^{200}$Hg. In $^{88}$Sr and $^{200}$Hg the combined data from above [30-31] and below [19, 25] $S_n$ are well described by $\Gamma_0=$ const; the widths $\Gamma_0 = 4.5$ and 3.3 MeV, respectively, correspond to $\Gamma(E_R) = 1.99$ MeV in Eq. (5) (to be discussed in the following). Even after smoothing the cross sections "pygmy" structures similar to the ones observed previously [4, 5] remain visible in these two spherical nuclei. Independent of that fact the data shown in Fig. 1 clearly discriminate against a spreading width $\Gamma_0$ which decreases with $E_\gamma^2$ as was asserted earlier [17, 32].

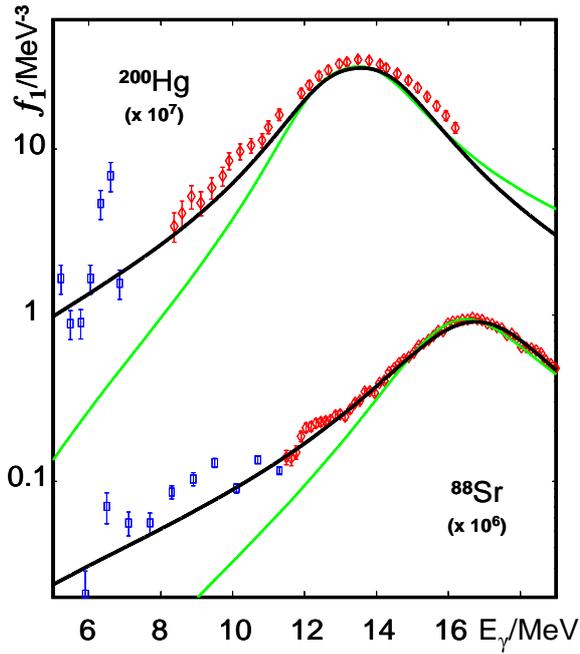

Fig. 1: Dipole strength functions $f_1(E_\gamma)$ for the spherical nuclei $^{200}$Hg (top: $^{nat}$Hg($\gamma$,n) [31]; $^{199}$Hg(n,$\gamma$) [25]) and $^{88}$Sr (bottom: $^{nat}$Sr($\gamma$,n) [30]; $^{88}$Sr($\gamma$,$\gamma$) [19]). The calculations are shown as thick lines for $\Gamma_0 =$ const. and for $\Gamma_0 \propto E_\gamma^2$ in thin. The data below $S_n$ are averaged over 250 and 600 keV respectively.

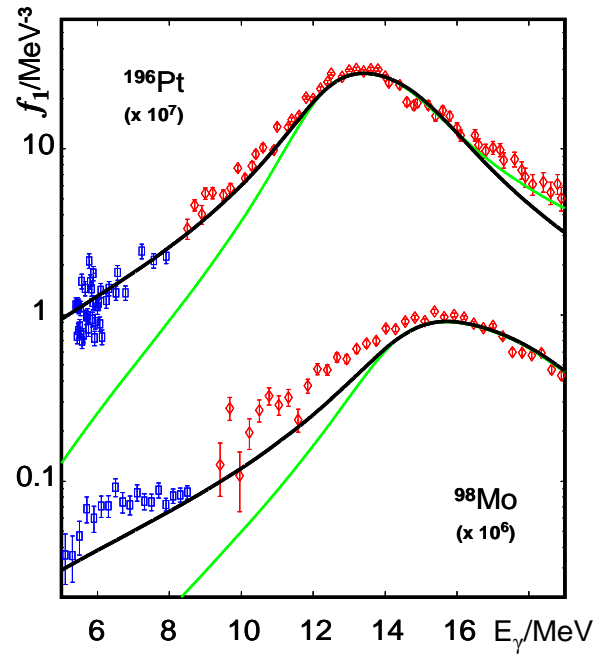

Fig. 3: Dipole strength functions $f_1(E_\gamma)$ for the triaxial nuclei $^{196}$Pt (top) and $^{98}$Mo (bottom). Above $S_n$ the data are from ($\gamma$,n)-experiments [40,41]; the data below $S_n$ are from ($\gamma$,$\gamma$) for $^{98}$Mo [20] and from ARC for $^{196}$Pt [26]. The thick lines are for $\Gamma_k =$ const., the thin lines represent $\Gamma_k \propto E_\gamma^2$.

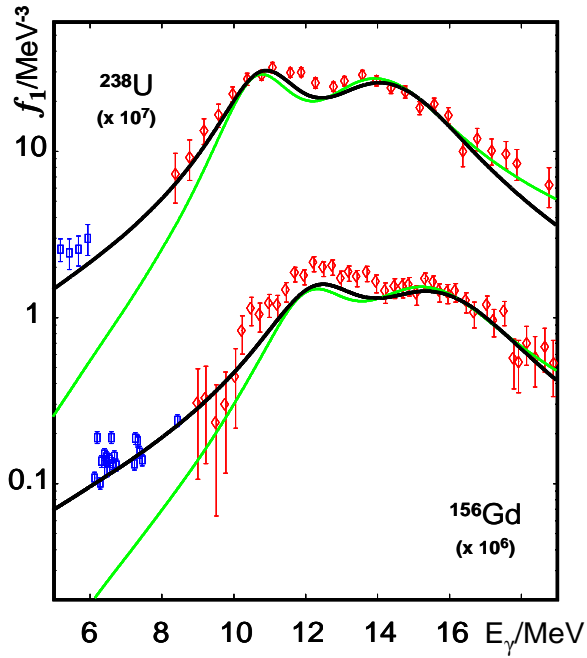

Fig. 2: Dipole strength functions $f_1(E_\gamma)$ from photon absorption [33] by the well deformed nuclei $^{238}$U (top) and $^{156}$Gd (bottom). The thick line depicts the parameterization as proposed here and the thin lines correspond to $\Gamma_k \propto E_\gamma^2$. The data below $S_n$ are from photon scattering [34] and ARC [26], respectively.

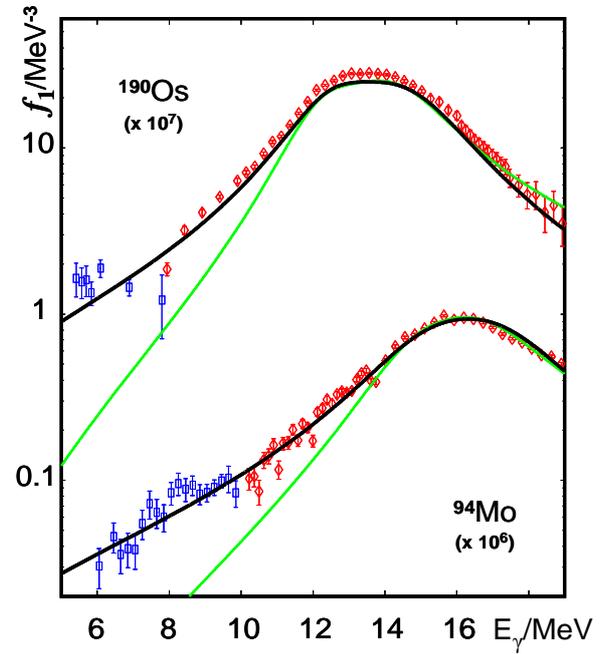

Fig. 4: Dipole strength functions $f_1(E_\gamma)$ for the transitional nuclei $^{190}$Os (top) and $^{94}$Mo (bottom). Above $S_n$ the data are from ($\gamma$,n)-experiments [44, 40]; the data below $S_n$ are from ($\gamma$,$\gamma$) for $^{94}$Mo [21] and obtained from properly normalized ARC [25] for $^{190}$Os [5]. The thick lines are for $\Gamma_k =$ const., the thin lines correspond to $\Gamma_k \propto E_\gamma^2$.

As mentioned above, fits with a single Lorentzian to the GDR data in nuclei with some deformation – even if it is small – do not yield a systematic A-dependence of the widths [14]. For all nuclei but the really spherical ones we now make use of the fact that the vibration frequency $E_k / \hbar$ along a given axis k is inversely proportional to the corresponding semi-axis length $R_k$. Using the Hill-Wheeler parameters β and γ of ellipsoidal shapes [9, 11] one obtains

$$E_k = E_0 \cdot R_0 / R_k = E_0 / \exp[\sqrt{5/4\pi} \cdot \beta \cos(\gamma - \tfrac{2}{3}k\pi)] \quad (3)$$

$R_0$ and $E_0$ are the radius and GDR energy of a spherical nucleus with the same mass A. Eq. (3) is easily derived for static quadrupole deformation, but due to the high frequency of the GDR oscillation it is assumed to also hold for the average deformations during quadrupole vibrations – in a quasi-adiabatic approximation. For the general case of a triaxial nucleus an incoherent sum of three Lorentz curves with widths $\Gamma_k$ is required [11, 20] which correspond to the dipole vibration of the nucleus along each of the principal axes k = 1, 2, 3:

$$\langle\sigma_\gamma(E_\gamma)\rangle = \frac{1.02 \cdot 11.9 \cdot Z \cdot N}{3 \cdot \pi \cdot A} \cdot \sum_{k=1}^{3} \frac{E_\gamma^2 \Gamma_k}{(E_k^2 - E_\gamma^2)^2 + E_\gamma^2 \Gamma_k^2} \quad (4)$$

with the photon energy $E_\gamma$, the resonance energies $E_k$ and the widths $\Gamma_k$ given in MeV and $\langle\sigma_\gamma(E_\gamma)\rangle$ in fm². The constant 1.02 in Eq. (4) accounts for the ratio of the Lorentz-based integral $\int \sigma_\gamma(E_\gamma) dE_\gamma$ to the value for Breit-Wigner curves. It was adjusted to hold for all $E_\gamma$, $E_k$, Z, A as dealt with in this paper (within < 2 %); the term 11.9 stems from the TRK-sum rule [9]. Multipolarities other than E1 are not included as we consider (4) as a reference which, when compared to measured data, allows to experimentally quantify all contributions, e.g. those due to velocity dependent or exchange forces. As mentioned, we make use of the well known fact that particle escape contributes to the width by a negligible amount [11].

For a parameterization of the spreading width we use results from hydrodynamical considerations [11] thereby adapting surface dissipation to the Goldhaber-Teller model of the GDR. For the widths of the different GDR components this results in a power law dependence on the respective resonance energy:

$$\Gamma_k(E_k) = \Gamma_0(E_0)\left(\frac{E_k}{E_0}\right)^\delta = 1.99 MeV \cdot \left(\frac{E_k}{10 MeV}\right)^\delta \quad (5)$$

This equation holds for the three modes k = 1,2,3 in a triaxial nucleus and for the exponent the value δ = 1.6 was derived from the one-body dissipation model [11]. We generalize this by using the proposed power law dependence for $\Gamma_0(E_0)$ as well, i.e. we apply it to different nuclei and thus reduce the number of parameters for the description of the GDR's. By that (2$^{nd}$ part of (5)) we relate the spreading width in all nuclei with mass A>80 to the respective resonance energy of their (up to three) GDR components.

From the very instructive compilation of apparent GDR widths in nuclei from Rb to U as obtained in (γ,n)-experiments at CEN Saclay [13, cf. Fig.5] it is obvious that the shape of the GDR peak is very strongly influenced by deformation-induced splitting. In axially deformed nuclei the GDR splits into two components and the higher (lower) energy one in prolate (oblate) nuclei should correspond to two times the absorption cross section as the other one.

Seemingly this is not observed experimentally [13]. By combining Eqs. (4) and (5) a larger width at the higher energy – and thus a reduced maximum – is predicted by our parameterization, even when the extra $E_\gamma$ in the denominator of Eq. (2) is considered.. We thus directly reproduce this aspect of the data for deformed nuclei. This is observed in Fig.2 for the two nuclei $^{156}$Gd and $^{238}$U; the data [33] shown for $\sigma_\gamma(E_\gamma)$ above $S_n$ were not obtained via (γ,n) but rather by observing photon absorption directly, a method which is free from ambiguities related to the detection of neutrons. For these nuclei an axial shape with β = 0.27 ( 0.29) was used in accordance to spectroscopic information [35] and the FRDM [7]. A good agreement to the data [26, 33-34] is observed only for $\Gamma_k =$ const($E_\gamma$). The large atomic charge of U required the subtraction of approximately 20% Delbrück scattering contribution [36] from the photon scattering data [33].

Various evidence has been presented pointing to the existence of nuclei which are triaxial in their ground state [37, 38], but only recently detailed experimental studies [22, 39] have determined accurate triaxiality parameters γ. To test our dipole-strength calculation for such nuclei we compare it in Fig. 3 to GDR data [40] of $^{98}$Mo, for which detailed Coulomb excitation studies [39] give γ = 23° and β = 0.18. For $^{196}$Pt [41] γ = 29° and β = 0.13 were used [41,42]. Like before, the calculation for $\Gamma_k =$ const($E_\gamma$) is satisfactory also in the threshold region and below.  As an especially sensitive test of our method we show in Fig. 4 the GDR in two transitional nuclei: $^{94}$Mo and $^{190}$Os. Spectroscopic investigations give β= –0.08 for $^{94}$Mo, β= –0.16 for $^{190}$Os [35] and γ=20°. With $\Gamma_k =$ const($E_\gamma$) a good description of the data is obtained. In the past GDR data for nuclei with small prolate or oblate deformation were often fitted by a single Lorentzian [14, 17] with the consequence that a seemingly large width is falsely attributed to spreading. Especially the width of Γ = 4.6 MeV at $E_0$ = 13.7 MeV as used [17] for $^{197}$Au is in clear disagreement to our Γ = 3.2 MeV at $E_0$ = 13.5 MeV for $^{200}$Hg and our average 3.3 MeV for $^{196}$Pt. This supports our view that only for clearly spherical nuclei a one-resonance fit is adequate.

When transitional and triaxial nuclei are considered to be quasi spherical one-resonance Lorentzian fits [13] to the GDR falsely result in too large widths. Thus it was proposed [17, 27] to use $\Gamma(E_\gamma) \propto E_\gamma^\alpha$ with α = 2 for such nuclei. This is an important ingredient of the so-called KMF-parameterization [32] which results in a reduced E1-strength in the low energy part of the GDR and below threshold. This ansatz was supported by the fact that in several "non-deformed" nuclei the dipole strength below threshold as derived from primary capture photons was low as compared to the Lorentzian. But, a photon energy dependence of the width was not indicated for the well deformed nuclei $^{157}$Gd [28] and $^{163}$Dy [23] and, in contrast, the necessary two-resonance fit resulted in such a small prediction for the strength at low energy that α = 0 could not be excluded by these data. As the Landau theory of Fermi liquids seemed to also justify α = 2 this problem of the proper α remained unsolved since many years. Previously, a new solution to this puzzle was searched for in a generalized Fermi liquid model [26]. By an addition developed to account for the quadrupole degrees of freedom the KMF-term [32] with α = 2 was complemented. For 8 nuclei between A=146 and A=198 the "traditional" Lorentz fits [14] were used. As we consider these misleading, we only use the raw experimental data. Very recently KMF – including α = 2 – was selected [43] as basis for a comprehensive "Compilation of giant electric dipole resonances..."

We propose not to rely on existing Lorentzian fits [14] and strive for an inclusion of shape degrees of freedom already in the analysis of the raw GDR data. This is at variance to previous work [44], in contrast to that we clearly separate the deformation induced widening from spreading effects already in the analysis of the raw data. We demonstrated in Figs.1-4 that for energies above 5 MeV an analysis of photon data accords to α ≈ 0 in heavy nuclei with A>80. We repeat here that for such nuclei the FRDM [7, 8] relates the centroid energy $E_0$ of the GDR to A and Z by only one additional parameter, the effective mass m*= 874 MeV/c². And by the use of the exponent δ =1.6 from hydrodynamics [11] and spectroscopic information [35, 39, 45] on the shape parameters only one new parameter is needed to describe the width of the GDR and its low energy tail. This parameter describes the width induced by the spreading of the E1-strength into the underlying quasi-continuum. From various test calculations we know that our findings are not sensitive to small changes in the FRDM or the shape parameters as used. Far above the GDR our few-parameter description may have shortcomings due to additional degrees of freedom coupling or competing to the GDR mode. And below 4 MeV the photon scattering data, e.g. those from the ELBE radiation laboratory [18-21], show a strong decrease of dipole strength as averaged over the few remaining spin 1 states. This subject is beyond the scope of the present work and may be confronted to Fermi liquid theory. At such low energy the influence of magnetic strength has to be studied [16, 22-23] and for large Z Delbrück scattering may become important [36].

The experimental cross section data can be compared to the calculation by point-wise adding the difference within the range studied. For the nuclei regarded here an agreement within a factor between 1.02 and 1.15 was found with an average of 1.08 (5). Even when considering the experimental uncertainties this good accordance to the TRK sum rule illustrates that the strength not coming from the GDR is small. In some of the eight nuclei resonance-like "pygmy" structures are seen below the GDR; their strength exceeds the expression proposed here for the smooth dipole strength by not more than a few % of the TRK sum.

In principle, a microscopic calculation may describe both, the more isolated "pygmy" strength as well as the quasi continuum characterized by close non-overlapping levels just below $S_n$ and overlapping resonances above. The random phase approximation (RPA) is known to treat complicated many body problems and it can be considered a density functional formalism when performed with a density dependent interaction. In the QRPA quadrupole phonons are included explicitly and respective calculations have been performed for various nuclei, mostly of spherical symmetry. Inspecting results from the only QRPA calculation available [6, 46] for all nuclei of this study we find that the parameterization as proposed here is closer to the data as compared to these QRPA predictions. They are presented in the RIPL-2 web-page [6] together with 6 different parameterizations for the dipole strength, including those mentioned above [14, 17, 26, 32]. As these are all based on the "traditional" Lorentzians [14], they are at variance to our findings and, despite additional parameters, they are inferior in their agreement to the combined data. The "challenge .. to find a .. model which can satisfactorily account for the data of both spherical and deformed nuclei" [26] is thus not met in RIPL-2.

Concluding: In most heavy nuclei the excursion from shell closure results in a widening of the GDR. Thus a two- or even three-component Lorentzian is required for the description of the energy dependent electric dipole strength. The deformation parameters have to be determined independently and accounted for in detail to extract the electric dipole spreading width. For all nuclei with A > 80 this spreading width is given by $\Gamma(E_k)=$ 1.99 MeV · $(E_k /10$ MeV$)^\delta$, with the exponent δ = 1.6 as derived from hydrodynamical considerations. This result of the GDR spreading only varying with resonance energy $E_k$ can be considered an indication of its common origin in all heavy nuclei. In contrast to this change with $E_k$ the dependence of the spreading on photon energy $E_\gamma$ is negligible, i.e. one may set α = 0 in an expansion $\Gamma(E_\gamma) \propto E_\gamma^\alpha$. This fact, of major importance at energies down to 5 MeV, is borne out by photon scattering as observed with bremsstrahlung [18-21] and by ARC data on low spin negative parity nuclei [25-26].

The (γ,n)-data as displayed in the figures are copied from the EXFOR-compilation [47]; they were, if applicable, adjusted in cross section according to a proposition made by Berman et al. [14].

**Acknowledgement**
Intensive discussions with F. Becvar, F. Dönau, S. Frauendorf, M. Krticka, P. Michel, K.D. Schilling and R. Wünsch are gratefully acknowledged.